\documentclass[10pt,aps]{revtex4}
\usepackage{amssymb}
\usepackage{latexsym}
\usepackage{epsfig}
\usepackage{float}
\usepackage{graphicx,epsfig, color}
\usepackage{graphicx}
\usepackage{subfigure}
\usepackage{hyperref}
\usepackage{amsmath}
\usepackage{xcolor}

\begin{document}
\title{\textbf{ Anisotropic surface tension and stability of quark matter modified by the vector interaction }}
\author{Yu-Ying He and Xin-Jian Wen  }

\affiliation{Institute of Theoretical Physics, State Key Laboratory
of Quantum Optics and Quantum Optics Devices, Shanxi University,
Taiyuan, Shanxi 030006, China}

\begin{abstract}
In this article, the surface tension and stability of quark matter modified by the vector interaction in a strong magnetic field are investigated in the quasiparticle model with the multiple reflection
expansion. The self-consistent thermodynamic treatment of the chemical-potential-dependent quark mass is maintained
by the effective bag function, which depends on both the chemical potential and the magnetic field. It is found that the vector interaction could enlarge the surface tension in both the parallel and transverse directions with respect to the magnetic field. In a stronger magnetic field region, the presence of the vector repulsive interaction leads to an increase in transverse surface tension with the magnetic field strength, which is opposite to the vanishing value without repulsive interaction in the previous work. Consequently, it is concluded that a moderate-intensity magnetic field is required for the formation of a quark matter bubble with the vector interaction. Finally, it is demonstrated that the vector interaction slightly reduces the stability of quark matter. 
\end{abstract}

%\pacs{04.70.-s, 04.30.-w, 04.50.-h, 04.20.Jb, 04.70.Bw, 04.70.Dy}

%\keywords{Quasi-topological gravity; Ads spacetime; Thermal stability. }

\maketitle

\section{Introduction}

Recently, numerous studies have focused on systems of strongly interacting matter under the influence of intense magnetic fields. It is well known that a strong magnetic field is relevant in the early universe, magnetars, and noncentral heavy ion collisions. It produces a significant effect on the equation of state in compact stars and the quantum chromodynamics (QCD) phase diagram. The magnetic field modifies the microscopic properties of quark matter with the corresponding macroscopic implications in compact stars \cite{Chakrabarty:1996te, Broderick:2000pe, PerezMartinez:2005av, Deb:2021ftm}. As pointed out in Refs. \cite{Ferrer:2010wz, Khalilov:2002rz}, in strong magnetic fields, the breaking of the {\it O}(3) rotational symmetry by the magnetic field results in the anisotropy of the total pressure, having a smaller value parallel than perpendicular to the field direction. In Ref. \cite{Lugones:2016ytl}, Lugones et al. pushed the investigation of the surface tension in longitudinal and transverse components with respect to the magnetic field in the bag model \cite{Lugones:2018qgu, Grunfeld:2020gnv}. The study of the surface tension of quark matter has attracted much attention because of its impact on heavy-ion collisions and astrophysics. Surface tension plays a key role in the understanding of neutron star (NS) interiors and of the most external layers of strange quark stars \cite{Xia:2016guv}. Surface tension is also an important aspect in the process of quark matter nucleation that leads to the formation of hybrid or strange quark stars because it determines the critical size and the nucleation time of the first quark-matter droplets \cite{Carmo:2013fr, Lugones:2011xv, Buballa:2012vm}.

In recent studies, researchers have extensively employed phenomenological quark models like the MIT bag model \cite{Sen:2022lig, Pal:2023quk, Chodos:1974je, Sen:2021cgl, Podder:2023dey} and the quark quasiparticle model \cite{Zhang:2021qh, Chu:2023rty, Ma:2023stj, Peng:1999gh, Wen:2005uf, Chu:2012rd, Benvenuto:1989kc} to investigate the thermodynamic properties of strange quark matter (SQM), quark stars, and hybrid stars. Several decades ago, the cold dense quark matter was investigated in the MIT bag model, which opened the door to the extensive research on quark matter\cite{Chodos:1974pn}. The interaction between quarks is represented by a phenomenological bag constant and produces a bag pressure on the non-interacting system \cite{Chodos:1974je, Farhi:1984qu}. The quark mass is considered to be infinite outside while it is constant within the bag. The quark quasiparticle model, as an extended bag model, has been developed in studying the bulk properties of dense quark matter at finite density and temperature. Taking advantage of the medium effect, the strong interactions between the elementary degrees of freedom are incorporated through the medium-dependent quasiparticle mass \cite{Ma:2023stj, Peshier:1994zf}. The significance of the quasiparticle model is the successful description of the confinement mechanism by the density and/or temperature-dependent bag function, via which the first-order deconfining phase transition was constructed and the critical end point was determined \cite{Srivastava:2010xa}. The medium effect in the quasiparticle model on the formation of the quark matter bubble will be of great interest. The aim of this work is to investigate the anisotropy of the surface tension of the bubble in two-flavor quark matter under strong magnetic fields with vector interactions, and to compare the results with cases where vector interactions are absent.

SQM is comprised of approximately equal numbers of $u$, $d$, and $s$ quarks. Based on various phenomenological models, it has long been suspected that SQM is the true ground state of strongly interacting systems \cite{Farhi:1984qu, Itoh:1970uw, Bodmer:1971we, Witten:1984rs}. Ever since Bodmer-Witten hypothesis suggested that clusters of SQM can be more stable than the most stable atomic nucleus, there has been a lot of interest in studying it. In this study, we will calculate the free energy per baryon for both two-flavor and three-flavor quark matter under strong magnetic fields, and investigate the impact of vector interactions on the stability of quark matter systems.

This paper is organized as follows. In Section \ref{sec:model}, we present the self-consistent thermodynamics of the magnetized quark matter in the quasiparticle model. In Section \ref{sec:result}, the numerical results for the confinement bag function and surface tension are shown in a strong magnetic field, as well as the free energy per baryon for two-flavor and three-flavor quark matter, and detailed discussions are focused on anisotropy of surface tension and the stability with vector interactions. The last section is a short summary.

\section{Thermodynamics of the quasiparticle model in strong magnetic fields}\label{sec:model}

In the quasiparticle model, the medium effect is taken
through a chemical-potential-dependent quark mass, while
the bag pressure is taken to be constant. In the dense
system, quarks interact with other quarks to create an
effective mass, which makes them act as quasiparticles\cite{Pal:2024nza}.
In the hard-dense loop (HDL) approximation, the effective quark propagator is generated by resumming one-loop self-energy diagrams, and then the effective quark mass is derived from the dispersion relation of this propagator in the zero-momentum limit\cite{Schertler:1996tq}. The effective quark mass $m_i^*$  can thus be expressed as\cite{Schertler:1996tq, Schertler:1997vv, Peshier:1999ww, Bannur:2007tk, Wen:2009zza}
\begin{equation}\label{mass}
m_i^*(\mu_i^*)=\frac{m_{i0}}{2}+\sqrt{\frac{m_{i0}^2}{4}+\frac{g^2{\mu_i^*}^{2}}{6\pi^2}},
\end{equation}
where $m_{i0}$ is the current mass of the quark and $g$ is the coupling constant. The repulsive effect of the quark interaction is also included here by introducing the vector meson as a mediator \cite{Pal:2023quk, Ju:2021nev}. The inclusion of a vector meson in the description of
quark matter leads to modifications of the chemical potential:
\begin{equation}
\mu_i^*=\mu_i-g_V V_0.
\end{equation}
Here $g_V$ is the vector interaction coupling and $m_V$ is the mass
of vector mesons. The nonzero mean field $V_0$ is calculated from the equation of motion for the vector meson,
\begin{equation}
m_V^2V_0=g_V\displaystyle\sum_i{n}_{i}.      \label{eq:gap}
\end{equation} 
where the equation of motion links the vector to the quark number density $n_i$, which will be given later. The vector field is not free but induced by the quarks through interactions. We can change the expression as
\begin{equation}
g_V V_0={G}_{V}\displaystyle\sum_i{n}_{i},
\end{equation}
where the dimensional vector interaction coupling constant is ${G}_{V}={\left ( \frac{{g}_{V}}{{m}_{V}}\right )}^{2}$.

In order to investigate the finite size effect of the bubble, we employ the
multiple reflection expansion (MRE)\cite{Farhi:1984qu, Balian:1970fw, Madsen:1993ka, Berger:1986ps}. In the MRE framework, the thermodynamic potential density of quark at zero temperature and in strong magnetic fields is\cite{Lugones:2021tee, He:2023gva},
\begin{eqnarray}\label{eq:5}
\Omega_{\mathrm{MRE},i} &=& \frac{d_i |q_iB_m|}{2\pi^2} \sum_{\nu=0}^{\nu_i^{\mathrm{max}}}
(2-\delta_{\nu 0})
  \int_{\Lambda_{\mathrm{IR}}}^{p_F}
    \left(
E_i-\mu_i^*
    \right)
\rho_\text{MRE}  \mbox{d}p_z-\frac{1}{2}{m}_{V}^{2}{V}_{0}^{2}\, ,
\end{eqnarray}
where the density of states in momentum space is \cite{Balian:1970fw, Berger:1986ps, Mardor:1991dt, Madsen:1994vp}
\begin{eqnarray}\label{eq:6}
\rho_{\text{MRE}}(\mu_i,p)=1+\frac{6\pi ^{2} }{pR}f_{S}+\frac{12\pi ^{2}}{\left ( pR\right )^{2}} f_{C}, 
\end{eqnarray}
with the functions
\begin{eqnarray}
f_{S}(\mu_i,p)=-\frac{1}{8\pi }\left( 1-\frac{2}{\pi }\arctan\frac{p}{m_{i}^{*} }\right ), 
\end{eqnarray}
and
\begin{eqnarray}
f_{C}(\mu_i,p)=\frac{1}{12\pi ^{2}}\left [ 1-\frac{3p}{2m_{i}^{*}}\left ( \frac{\pi }{2}-\arctan \frac{p}{m_{i}^{*}}\right )\right ].
\end{eqnarray}
The single particle energy eigenvalue is $E_i=\sqrt{{m_i^*}^2+p_z^2+ 2 \nu_i
	|q_i B_m|}$ , and the upper limit $\nu_i^{\mathrm{max}}$ of the summation
index $\nu_i$ is $\nu_i^{\mathrm{max}}=\frac{{\mu_i^*}^2-{m_i^*}^2}{2 |q_i B_m|}$.
Generally, the modification of the density of states
in the MRE framework constrains the low limit on the infrared cutoff
due to the fact that $\rho_\text{MRE}$ becomes negative at small
momenta \cite{Kiriyama:2005eh}. To obtain the value of $\Lambda_\mathrm{IR}$, we have to solve the following equation numerically
\begin{equation} \label{eq:9}
\rho_{\text{MRE}}(\mu_i,p)=0,  
\end{equation}
with respect to the momentum and take the larger root as the infrared cutoff $\Lambda_\mathrm{IR}$.

The total thermodynamic potential including electrons in the framework of the
quasiparticle model is written as
\begin{eqnarray}\label{eq:10}
\Omega=\sum_i \Omega_{\mathrm{MRE},i}+\Omega_e+B_0,
\end{eqnarray}
where the thermodynamic potential of electrons is:
\begin{eqnarray}
\Omega_e=\frac{|eB_m|}{2\pi^2} \sum_{\nu=0}^{\nu_i^{max}}
(2-\delta_{\nu 0})
\int_{0}^{p_F}
(E-\mu_e)\mbox{d}p_z.
\end{eqnarray}
The quark number density $n_i$ can be derived from the full derivative of the thermodynamic potential density $\Omega(\mu_i,m_i)$.
\begin{eqnarray}
n_i=-\frac{d\Omega}{d\mu_i}=-\sum_i\bigg(\frac{\partial \Omega_{\mathrm{MRE},i}}{\partial\mu_i}+\underbrace{\frac{\partial \Omega_{\mathrm{MRE},i}}{\partial m_i^*}\frac{\partial m_i^*}{\partial \mu_i}+\frac{\partial B_i^*}{\partial m_i^*}\frac{\partial m_i^*}{\partial \mu_i}}_0 \bigg).\label{eq:12}
\end{eqnarray}
As in the previous approach, we require the last two terms to be cancelled out. Then the medium-dependent bag function is expressed as
\begin{eqnarray}
B_i^*(\mu_i^*)=-\frac{d_i |q_iB_m|}{2\pi^2}
\sum_{\nu=0}^{\nu_i^{\mathrm{max}}} (2-\delta_{\nu 0})\int_{m_i^*}^{\mu_i^*} \int_{\Lambda_{\mathrm{IR}}}^{p_F}
\left \{ \frac{m_i^*}{E_i}\rho_\text{MRE}-\frac{3}{2R}\frac{E_i-\mu_i^*}{{E_i}^2}\right. \nonumber \\
\left.+\frac{3\left (E_i-\mu_i^* \right )}{2{m_i^*}^{2}{R}^{2}}\left ( -\frac{m_i^*}{{E_i}^2}+\frac{arctan(\frac{m_i^*}{p})}{p}\right ) \right \}\frac{dm_i^*}{d\mu_i^*}dp_zd\mu_i^*.\label{eq:13}
\end{eqnarray}
In the evaluation of the first term of Eq.(\ref{eq:12}), the single particle energy $E_i(p)$ and density of states $\rho_\mathrm{MRE}$$(\mu_i,p)$ are represented in the form of a function with the arguments. So,
\begin{eqnarray}
\frac{\partial \Omega_i}{\partial \mu_i}&=&-\frac{d_i|q_i B_m|}{2\pi^2}\Bigg\{ \int_{\Lambda_{\mathrm{IR}}}^{p_F}\rho_{\mathrm{MRE}}(\mu_i,p)dp_z+\underbrace{[E_i(p_F)-\mu_i]}_0 \rho_{\mathrm{MRE}}(\mu_i,p_F) \frac{dp_F}{d\mu_i}\nonumber \\
&& \phantom{-\frac{d_i|q_i B_m|}{2\pi^2}} - [E_i(\Lambda_{\mathrm{IR}})-\mu_i] \underbrace{\rho_{\mathrm{MRE}}(\mu_i,\Lambda_{\mathrm{IR}})}_0\frac{d\Lambda_{\mathrm{IR}}}{d\mu_i} \Bigg\},
\end{eqnarray}
where the second term is from the derivative of the upper limit of the integral multiplied by the value of the integrand at that upper limit, and it vanishes due to the definition of Fermi energy. The third term corresponds to the contribution from the lower limit, which is zero by virtue of condition (\ref{eq:9}). So the quark number density $n_i$ can be derived from the contribution of the first term,
\begin{equation}
{n}_{i}=\frac{d_i |q_iB_m|}{2\pi^2}
\sum_{\nu=0}^{\nu_i^{\mathrm{max}}} (2-\delta_{\nu 0})\int_{\Lambda_{\mathrm{IR}}}^{p_F}
\rho_\text{MRE}  \mbox{d}p_z.
\end{equation}
 At this point, we have already proven the thermodynamic consistency including the effect of the $\mu$-dependent IR cutoff. In the Appendix we discuss the steps to demonstrate the thermodynamic consistency in the framework of effective mass and the MRE formulation.
 
Hereafter, we refer to the thermodynamic potential $\Omega_{\mathrm{MRE},i}$ as $\Omega_{\mathrm{MRE},i}^\parallel$ in order to study the anisotropy. According to the power dependence on the radius, the thermodynamic potential density can be decomposed into volume, surface, and curvature terms. Similar to the result in Ref. \cite{Lugones:2021tee}, the contribution of $i$-flavor can be expressed as
\begin{eqnarray}
\Omega_{\mathrm{MRE},i}^\parallel =\Pi^\parallel_i  +\sigma_i^\parallel\frac{S}{V}+\gamma_i^\parallel\frac{C}{V}. \label{omegapara}
\end{eqnarray} 
For a sphere of radius $R$, we denote the bulk volume  $V=4 \pi R^3/3$, the surface area $S=4\pi R^2$, and the integrated mean curvature $C=8\pi R$. By comparing the power dependence on the radius on both sides of Eq.(\ref{omegapara}), we can extract the terms independent of $R$ for the bulk density $\Pi^\parallel$, the surface term $\sigma^\parallel$ and the curvature term $\gamma^\parallel$ from the dependence of $1/R$ and $1/R^2$ respectively. According to the expression of the thermodynamic potential in Eq.(\ref{eq:10}), each density will be regarded as a sum of two parts, $\Pi_i^\parallel=\Omega_{V,i}^\parallel+B_{V,i}^\parallel$. The quasiparticle contribution $\Omega_{V,i}^\parallel$ and the vacuum contribution $B_{V,i}^\parallel$ are written as
\begin{equation}
\Omega_{V,i}^\parallel=\frac{d_i |q_iB_m|}{2\pi^2} \sum_{\nu=0}^{\nu_i^{\mathrm{max}}}
(2-\delta_{\nu 0})
\int_{\Lambda_{\mathrm{IR}}}^{p_F}
\left(
E_i-\mu_i^*
\right)  \mbox{d}p_z+{\Omega }_{V}^{0}
\end{equation}
with the contribution of meson interaction
\begin{equation}
{\Omega }_{V}^{0}=-\frac{1}{2}{G}_{V}\displaystyle\sum_{i=u,d}\left[\frac{d_i |q_iB_m|}{2\pi^2}
\sum_{\nu=0}^{\nu_i^{\mathrm{max}}} (2-\delta_{\nu 0})\int_{\Lambda_{\mathrm{IR}}}^{p_F}{d}p_z\right]^2
\end{equation}
\begin{equation}
B_{V,i}^\parallel=-\frac{d_i |q_iB_m|}{2\pi^2} \sum_{\nu=0}^{\nu_i^{\mathrm{max}}}
(2-\delta_{\nu 0})
\int_{m_i*}^{\mu^*}\int_{\Lambda_{\mathrm{IR}}}^{p_F}
\frac{m_i^*}{E_i}
\mbox{d}p_z \mbox{d}\mu_i^*
\end{equation}

Correspondingly, the second term in Eq. (\ref{omegapara}) represents the surface tension and can be expressed as a sum of two parts:
\begin{equation}
\sigma_i^\parallel=\Omega_{S,i}^\parallel+B_{S,i}^\parallel. 
\end{equation}
The quasiparticle contribution can be integrated by the following formula
\begin{eqnarray}
\Omega_{S,i}^\parallel=-\frac{d_i |q_iB_m|}{4\pi^2} \sum_{\nu=0}^{\nu_i^{\mathrm{max}}}
(2-\delta_{\nu 0})  \int_{\Lambda_{\mathrm{IR}}}^{p_F}  \frac{E_i-\mu_i^*}{p} \arctan(\frac{m_i^*}{p}) \mbox{d}p_z+{\Omega }_{S}^{0},
\end{eqnarray}
where
\begin{eqnarray}
{\Omega }_{S}^{0}&=&-{G}_{V}\displaystyle\sum_{i,j=u,d}\left[\frac{d_i |q_iB_m|}{2\pi^2}
\sum_{\nu=0}^{\nu_i^{\mathrm{max}}} (2-\delta_{\nu 0})\int_{\Lambda_{\mathrm{IR}}}^{p_F}{d}p_z\right. \nonumber \\
&\times& \left.\frac{d_j |q_jB_m|}{4\pi^2}\sum_{\nu=0}^{\nu_j^{\mathrm{max}}} (2-\delta_{\nu 0})\int_{\Lambda_{\mathrm{IR}}}^{p_F}\left(-\frac{1}{p}\mbox{arctan}(\frac{m_j^*}{p})\right){d}p_z \right]. \label{eq:22}
\end{eqnarray}
The vacuum contributes to the surface tension as
\begin{eqnarray}
B_{S,i}^\parallel = \frac{d_i |q_iB_m|}{4\pi^2} \sum_{\nu=0}^{\nu_i^{\mathrm{max}}}
(2-\delta_{\nu 0}) \int_{m_i^*}^{\mu_i^*} \int_{\Lambda_{\mathrm{IR}}}^{p_F}
\left[ \frac{E_i-\mu_i^*}{E_i^2} +\frac{m_i^*}{E_i
	p}\arctan(\frac{m_i^*}{p})\right]\frac{dm_i^*}{d\mu_i^*} \mbox{d}p_z \mbox{d}\mu_i.
\end{eqnarray}

In literature, the transverse pressure is usually defined as
$P^\perp=-\Omega^\perp=P^\parallel-\mathcal{M} B_m$, where the magnetization
susceptibility $\mathcal{M} $ can be derived by the relation $\mathcal{M}=-\frac{\partial\Omega^\parallel}{\partial B_m}$ with $P^\parallel = -\Omega^\parallel$.
The transverse thermodynamic potential density is related to the parallel term by the magnetization
\begin{equation} \Omega^\perp=\Omega^\parallel+ \mathcal{M}B_m. \label{Mag}
\end{equation}
Just like the approach of the parallel term (\ref{omegapara}), the transverse term is also divided into three parts according to the power dependence on the radius,
\begin{eqnarray}
\Omega_{\mathrm{MRE},i}^\perp=\Pi^\perp_i +\sigma_i^\perp \frac{S}{V}+\gamma_i^\perp\frac{C}{V}. 
\end{eqnarray} 
The bulk contribution $\Pi_i^\perp$ is a sum of quasiparticle contribution $\Omega_{V,i}^\perp$ and the vacuum contribution $B_{V,i}^\perp$, which can be evaluated as
\begin{equation}
\Omega_{V,i}^\perp=-\frac{{d_i |q_iB_m|}^2}{2\pi^2}\sum_{\nu=1}^{\nu_i^{\mathrm{max}}}\nu \int_{\Lambda_{\mathrm{IR}}}^{p_F}\frac{\mbox{d}p_z}{E_i}+{{\Omega }_{V}^{0}},
\end{equation}
\begin{equation}
B_{V,i}^\perp=\frac{{d_i |q_iB_m|}^2}{2\pi^2}\sum_{\nu=0}^{\nu_i^{\mathrm{max}}}\nu\int_{m_i^*}^{\mu_i^*} \int_{\Lambda_{\mathrm{IR}}}^{p_F}\frac{m_i^*}{{E}_{i}^{3}}\frac{dm_i^*}{d\mu_i^*} \mbox{d}p_z \mbox{d}\mu_i.
\end{equation}
The transverse surface term is $\sigma_i^\perp=\Omega_{V,i}^\perp+B_{S,i}^\perp$, which can be evaluated respectively as
\begin{eqnarray}
{\Omega_{S,i}^\perp}=\frac{{d_i |q_iB_m|}^2}{2\pi^2}\sum_{\nu=1}^{\nu_i^{\mathrm{max}}}\nu \int_{\Lambda_{\mathrm{IR}}}^{p_F}\frac{1}{p}\arctan(\frac{m_i^*}{p})\frac{\mbox{d}p_z}{E_i}+{{\Omega }_{S}^{0}},
\end{eqnarray}
\begin{eqnarray}
{B_{S,i}^\perp}=\frac{{d_i |q_iB_m|}^2}{2\pi^2}\sum_{\nu=0}^{\nu_i^{\mathrm{max}}}\nu\int_{m_i^*}^{\mu_i^*} \int_{\Lambda_{\mathrm{IR}}}^{p_F}\left [ \frac{1}{{p}^{2}E_i}-2\frac{{\mu}_{i}^{*}}{{E}_{i}^{4}}+\frac{m_i^*}{{p}^{3}{E}_{i}^{3}}\left ({E}_{i}^{2}+{p}^{2} \right )\arctan(\frac{m_i^*}{p})\right ]\frac{dm_i^*}{d\mu_i^*} \mbox{d}p_z \mbox{d}\mu_i.
\end{eqnarray}
Although the magnetic energy density becomes relevant at ${10}^{14}$ G \cite{Terrero:2017mxc}, the pure electromagnetic Maxwell term ${B_m^2}/{2}$ is independent of the physical properties of the thermal QCD medium and plays no role in the thermodynamics of the system.  It only enters into the renormalization prescription \cite{Jia:2014ffa}. In our present consideration, the contribution of the pure magnetic field term is the same inside and outside the sphere, so it will not have any effect on the surface tension of the finite-size droplet.

The mixed phase contains quark matter droplets in
chemical equilibrium under weak interactions, which
means that the chemical potentials of different species
($u$, $d$, $s$ quarks and $e$) are related by
\begin{eqnarray}
\mu_s=\mu_d=\mu_u+\mu_e.
\end{eqnarray}
The charge neutrality condition reads
\begin{eqnarray}
\frac{2}{3}{n}_{u}-\frac{1}{3}{n}_{d}-\frac{1}{3}{n}_{s}-{n}_{e}=0,
\end{eqnarray}
where the electron number density is
\begin{eqnarray}
{n}_{e} = \frac{|eB_m|}{2\pi^2}
\sum_{\nu=0}^{\nu_i^{\mathrm{max}}} (2-\delta_{\nu 0})\int_{0}^{p_F}
{d}p_z.
\end{eqnarray}

\section{Numerical results and conclusion}\label{sec:result}
In the present paper, we investigate the anisotropy of surface tension in two-flavor quark matter with vector interactions in Section A, comparing the results with the case without vector interactions, and subsequently calculate the free energy per baryon for both two-flavor and three-flavor quark matter in Section B. We adopt the current quark masses $m_{u0}$=$m_{d0}$=$m_{0}$=5.6 MeV for up and down quarks, and $m_{s0}=135.7$ MeV for strange quarks. For both two-flavor and three-flavor quark matter, the bag constant is taken as ${B}_{0}$=(145 MeV)$^4$.

\subsection{Anisotropic surface tension}
The chemical-potential-dependent bag function plays an important role in the description of the deconfinement transition\cite{He:2023gva}. In Fig. 1, we investigate the behavior of the bag function as a function of the chemical potential, taking into account the effects of the magnetic field, vector interaction, and coupling constant. The results show that the bag function decreases as the chemical potential increases, indicating the occurrence of deconfinement. Additionally, when the coupling constant increases and/or the magnetic field becomes stronger\cite{He:2023gva}, the bag function also decreases. In contrast, compared to the case without vector interaction, the introduction of vector interaction increases the bag function. These behaviors can be explained by Fig. 2, where we investigate the dependence of the infrared cutoff on the chemical potential with and without vector interactions under different magnetic field strengths and coupling constants. According to Eq. (\ref{eq:13}), it can be inferred that an increasing infrared cutoff leads to a progressive reduction of the inner integration domain, consequently causing a monotonic decrease in the bag function. In addition, we show in Fig. 2 the fixed infrared cutoff (horizontal dotted line) employed in the MIT bag model, which is obtained by setting the mass in Eq. (\ref{eq:9}) to a constant value of $m_0$, for comparison with the medium-dependent infrared cutoff. It is evident that the infrared cutoff in the MIT bag model is constant and always lower than the medium-dependent one. This discrepancy arises as a direct consequence of our treatment of the effective mass within the modified MRE framework.
\begin{figure}[H]
	\centering
	\begin{minipage}{10cm}
		\includegraphics[width=10cm]{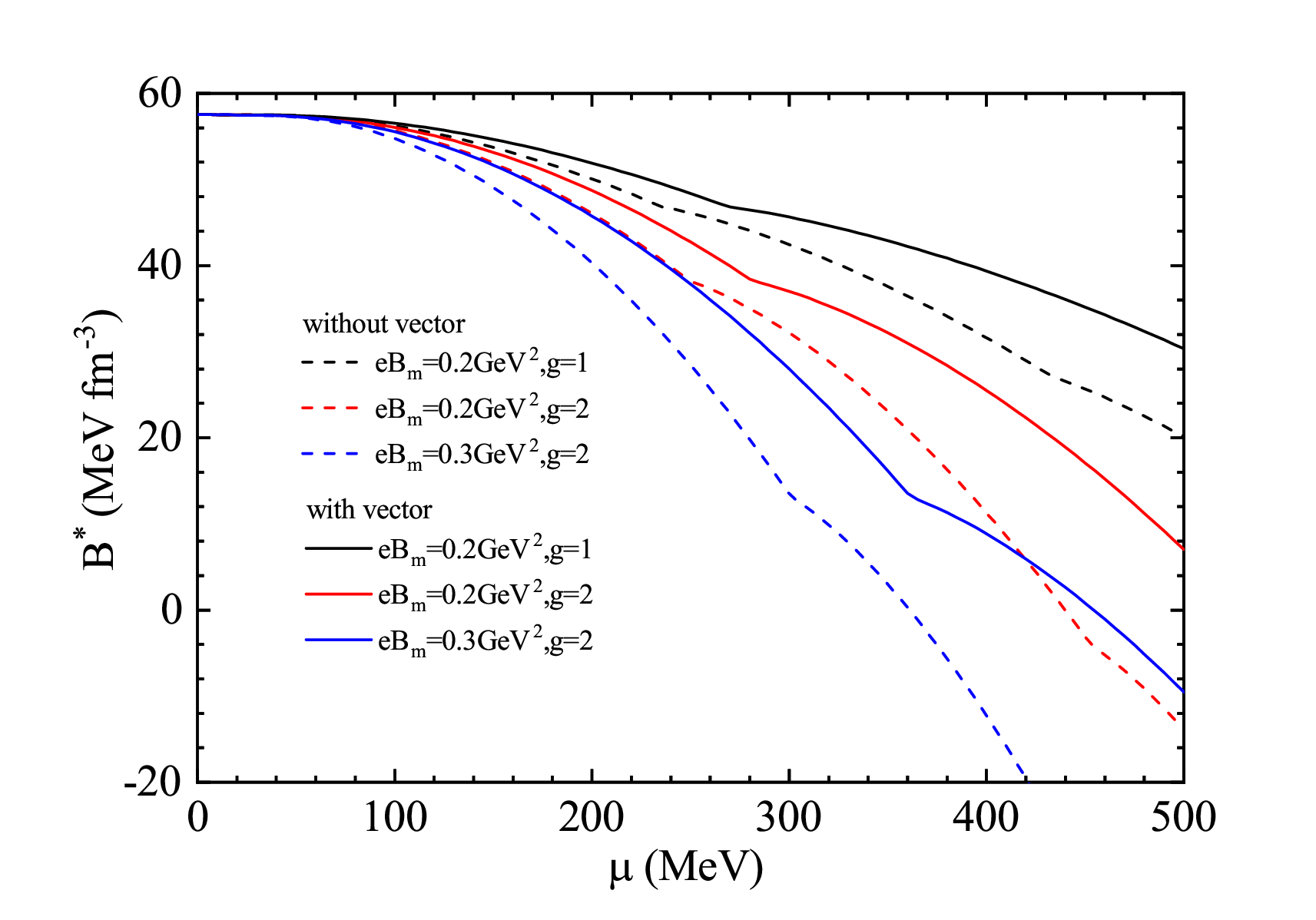}% Here is how to import EPS art
	\end{minipage}
	\caption{\label{fig:wide}The effective bag function $B^*(\mu_i^*)$ is shown as a function of the chemical potential for different coupling constants and magnetic fields, both with and without vector interaction. The dashed line represents the case without vector interaction, while the solid line represents the case with vector interaction.}\label{fig:1}
\end{figure}
\begin{figure}[H]
	\centering
	\begin{minipage}[b]{0.48\linewidth}
		\includegraphics[width=\linewidth,height=6cm,keepaspectratio]{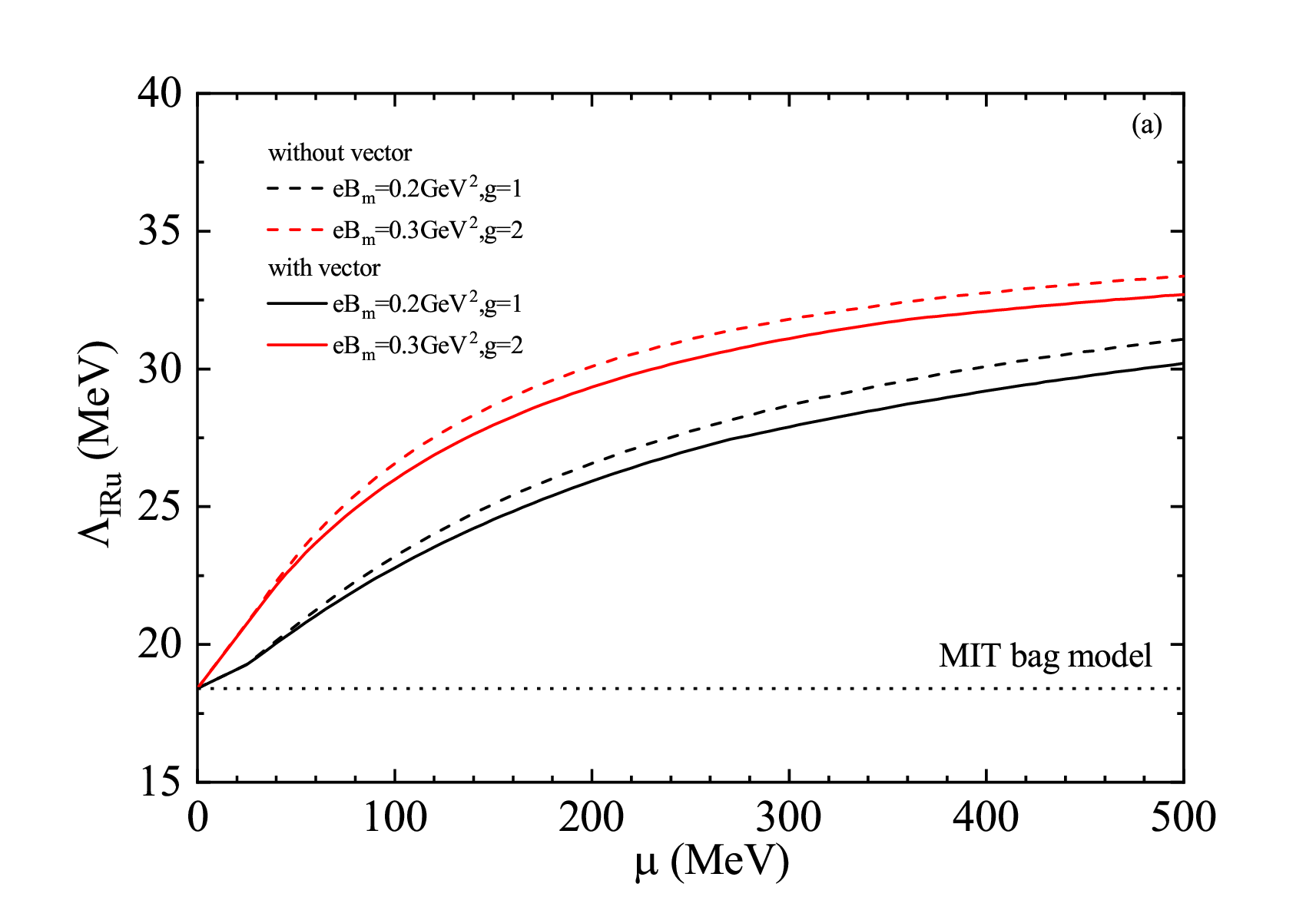}
	\end{minipage}
	\hfill
	\begin{minipage}[b]{0.48\linewidth}
		\includegraphics[width=\linewidth,height=6cm,keepaspectratio]{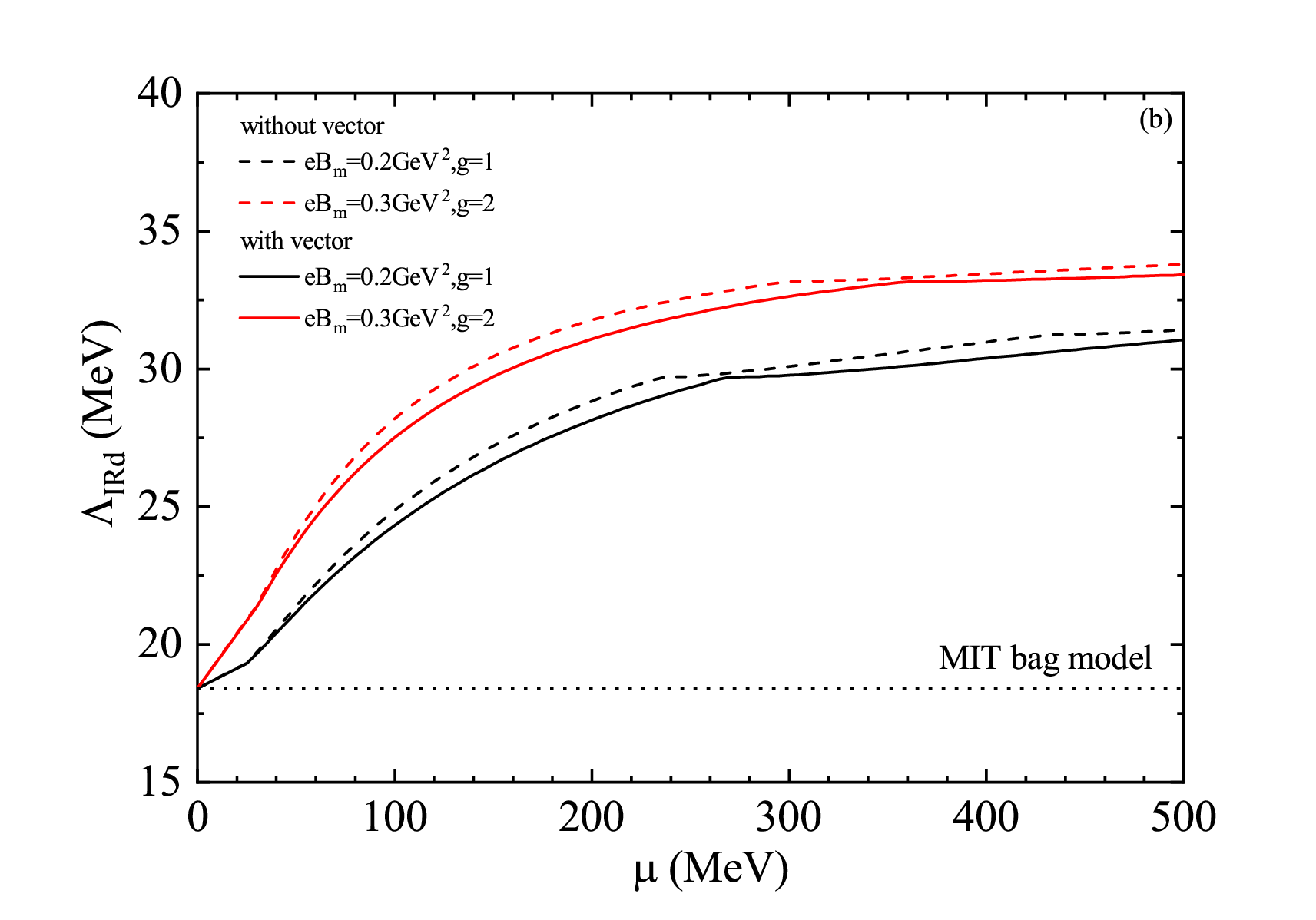}
	\end{minipage}
	\caption{\label{fig:wide}Dependence of the infrared cutoff on chemical potential for systems with (solid curves) and without (dashed curves) vector interaction, under various magnetic field strengths and coupling constants.}\label{fig:2}
\end{figure}
In Fig. 3, we investigate the effect of finite volume on the bag function with and without vector interaction. The dashed line represents the case without vector interaction, while the solid line corresponds to the case with vector interaction. The four horizontal lines are the corresponding bag functions for bulk SQM\cite{He:2023gva}. The bag function decreases monotonically with increasing radius, asymptotically approaching the constant value $B_0$ as $R\rightarrow \infty $, demonstrating the  disappearance of finite-size effects. Additionally, we compare the cases with and without vector interactions, finding that the bag function in the presence of vector interactions is larger than that in their absence. Consideration of the curvature term leads to a decrease in the bag function.
\begin{figure}[H]
	\centering
	\begin{minipage}{10cm}
		\includegraphics[width=10cm]{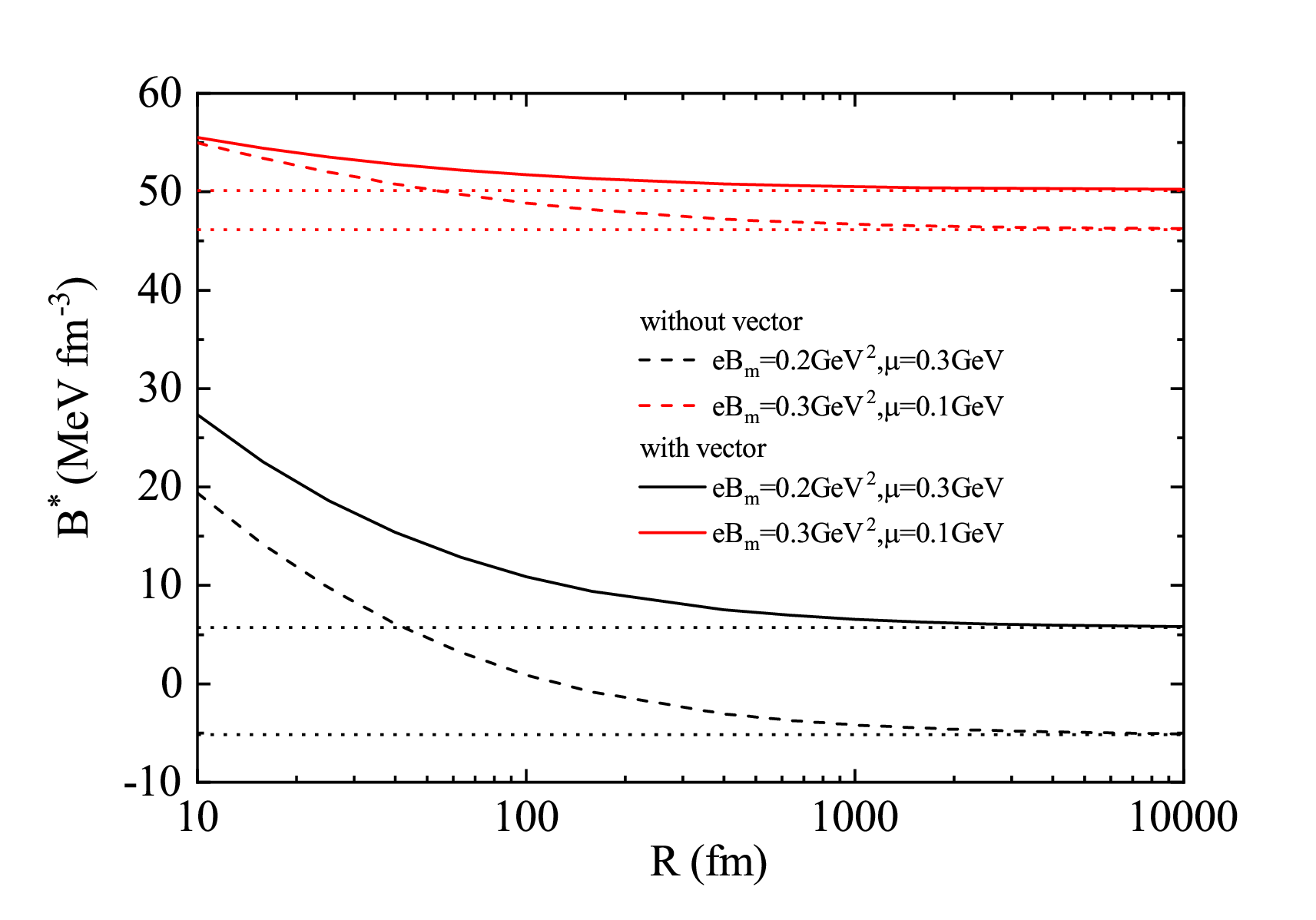}% Here is how to import EPS art
	\end{minipage}
	\caption{\label{fig:wide}The effective bag function $B^*(\mu_i^*)$ is shown as a function of the radius of the spherical system. The dashed line represents the case without vector interaction, whereas the solid line corresponds to the case with vector interaction. }\label{fig:3}
\end{figure}
To investigate the role of geometric effects in the thermodynamics of finite-size systems, we analyze the behavior of the thermodynamic potential and its decomposition into volume, surface, and curvature contributions. In Fig. 4(a), we present the thermodynamic potential $\Omega$ as a function of the vector coupling constant $G_V$, calculated under the conditions of radius $R$=10 fm, chemical potential $\mu$=100 MeV, and magnetic field strength $eB_m$=0.3 GeV$^2$. The results show that $\Omega$ increases monotonically with $G_V$. This behavior reflects the repulsive nature of the vector interaction: within the effective theory framework,  $G_V$ characterizes the strength of repulsion among fermions, and its increase systematically raises the free energy of the system.
Fig. 4(b) compares the behavior of the volume term $\Omega_V$, surface term $\Omega_S$, and curvature term $\Omega_C$ in the thermodynamic potential as functions of the chemical potential 
$\mu$, for a fixed vector coupling constant ${G}_{V}$=0.2 fm$^{2}$. In terms of magnitude, the three contributions consistently satisfy $\Omega_V>\Omega_S>\Omega_C$, indicating that the thermodynamic behavior of the system is dominated by bulk properties, with surface and curvature effects appearing as first-order and second-order geometric corrections, respectively, exhibiting a clear pattern of successive suppression. As the chemical potential $\mu$ increases, the three terms display distinctly opposite trends: the volume term $\Omega_V$ and the curvature term $\Omega_C$ decrease monotonically, while the surface term $\Omega_S$ increases monotonically.
\begin{figure}[H]
	\centering
	\begin{minipage}[b]{0.48\linewidth}
		\includegraphics[width=\linewidth,height=6cm,keepaspectratio]{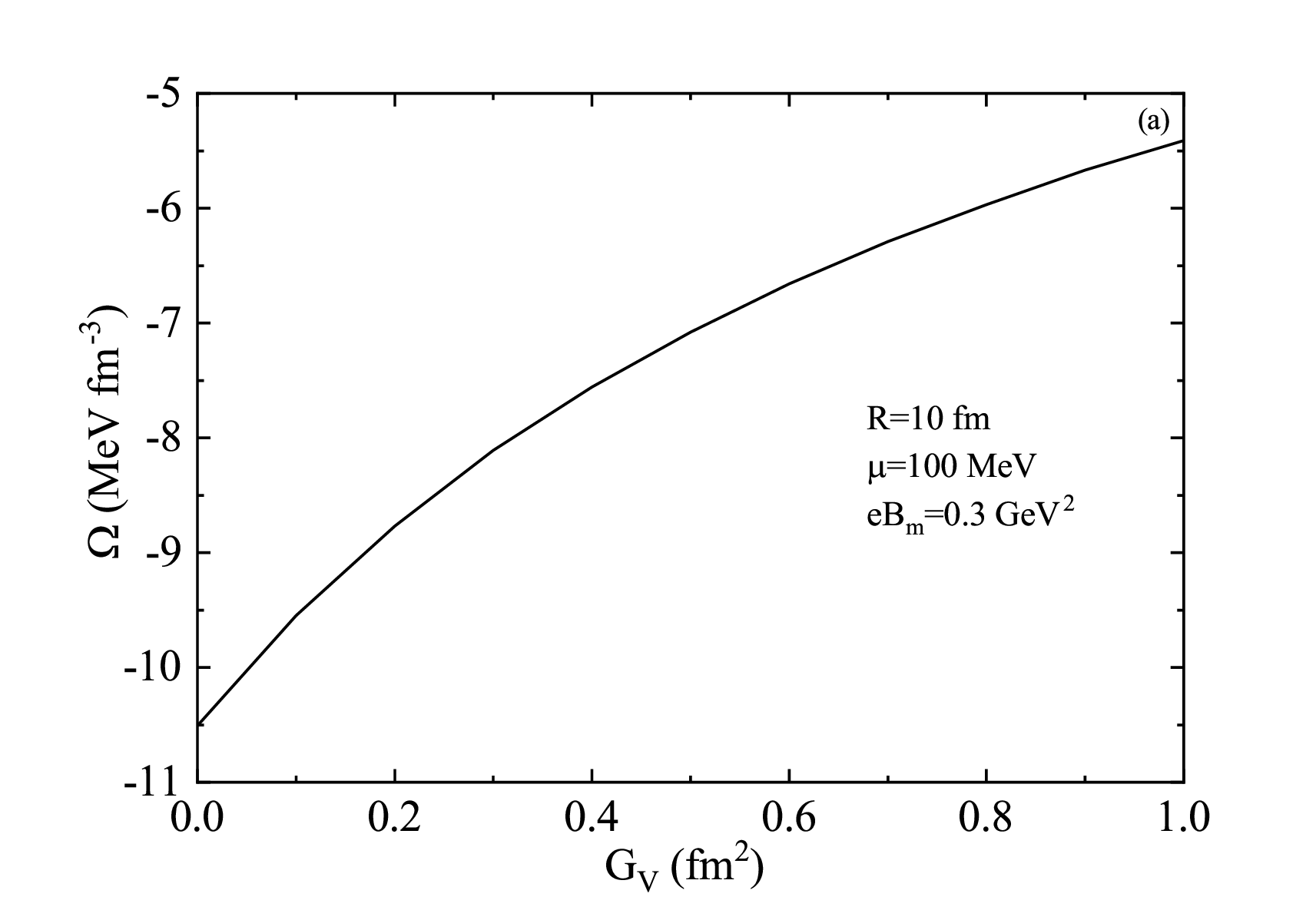}
	\end{minipage}
	\hfill
	\begin{minipage}[b]{0.48\linewidth}
		\includegraphics[width=\linewidth,height=6cm,keepaspectratio]{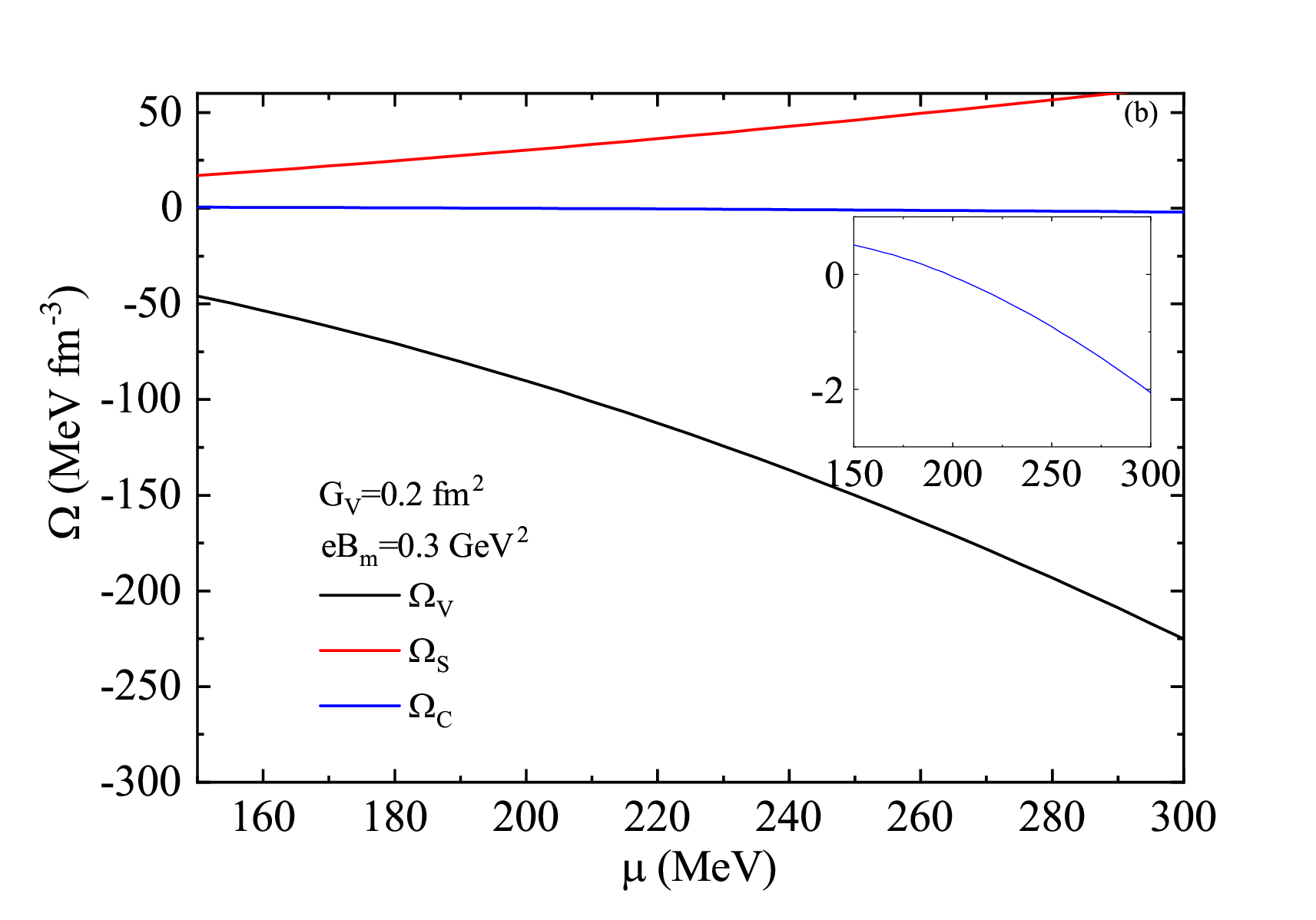}
	\end{minipage}
	\caption{\label{fig:wide}(a) The variation of the thermodynamic potential $\Omega$ with the vector coupling constant ${G}_{V}$ under the conditions of radius $R$=10 fm, chemical potential $\mu$=100 MeV, and magnetic field strength $eB_m$=0.3 GeV$^2$. (b) The variation of the volume term $\Omega_V$, surface term $\Omega_S$, and curvature term $\Omega_C$ of the thermodynamic potential with the chemical potential $\mu$ at a fixed vector coupling constant ${G}_{V}$=0.2 fm$^{2}$.}\label{fig:4}
\end{figure}
The surface tension is characteristic of the two-phase system. For example, the amount of the surface tension between the liquid drop and the gas phase is dominant in the raindrop formation\cite{He:2023gva}. At larger chemical potential, a direct transition from the vacuum to quark matter happens possibly depending on the surface tension of bubble quark matter\cite{Fraga:2018cvr}. Fig. 5 shows the surface tension as a function of magnetic field under the conditions of $g$=3, $R$=10 fm, and $\mu$=250 MeV. The colorful curves show the results of the quasiparticle model $G_v$=0, 0.1, 0.2, 0.3, 0.4 fm$^2$ and the MIT bag model (${G}_{V}$=0). The solid lines represent the surface tension components parallel to the magnetic field direction, while the dashed lines indicate the transverse components. It is shown that the surface tension in the quasiparticle model is significantly higher than that in the MIT bag model. The medium effects markedly enhance the surface tension. Furthermore, within a certain range of magnetic field strength, the parallel surface tension increases with the magnetic field strength $eB_m$, while the transverse surface tension shows a decreasing trend with increasing magnetic field. Meanwhile, the introduction of vector interactions enhances both the parallel and transverse surface tensions, indicating that vector interactions enhance the overall rigidity of the system and consequently increase the surface tension. In our present work, the curvature effect as a higher-order term enhances the surface tension.
\begin{figure}[H]
	\centering
	\begin{minipage}{10cm}
		\includegraphics[width=10cm]{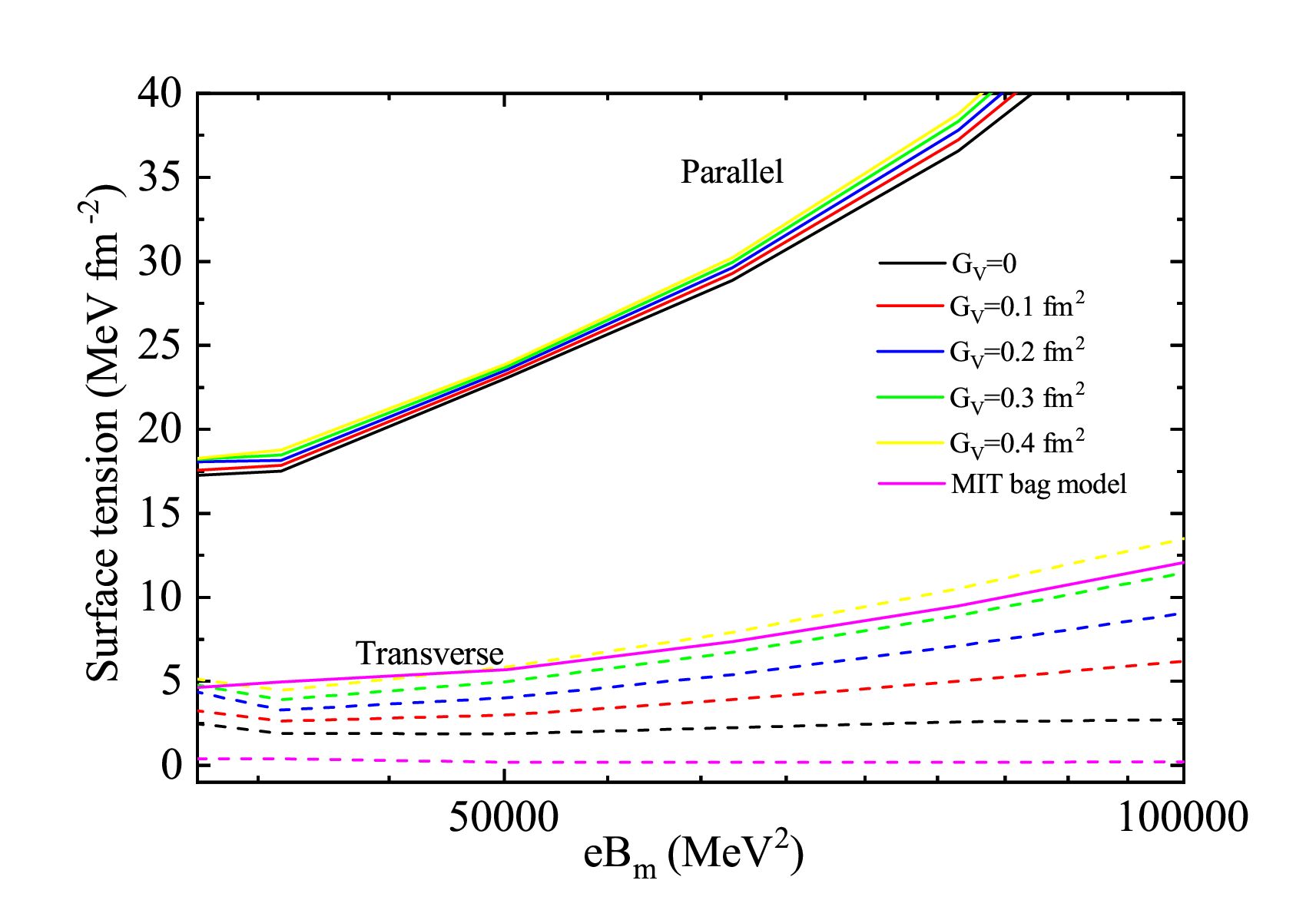}% Here is how to import EPS art
	\end{minipage}
	\caption{\label{fig:wide}The parallel and transverse surface tensions are plotted as functions of the magnetic field strength at zero temperature. The solid line represents the parallel surface tension, whereas the dashed line corresponds to the transverse surface tension. }\label{fig:5}
\end{figure}
To more intuitively illustrate the influence of vector interactions on the surface tension, Fig. 6 displays the dependence of the parallel and transverse surface tensions on the vector coupling constant under the same parameter conditions ($g$=3, $R$=10 fm, and $\mu$=250 MeV) and a fixed magnetic field strength of $eB_m$=0.05 GeV$^2$, for both the quasiparticle model (with a chemical-potential-dependent mass) and the MIT bag model (with a fixed mass). The solid lines represent the parallel surface tensions, while the dashed lines correspond to the transverse surface tensions. For both models, the parallel and transverse surface tensions increase monotonically with the vector coupling constant, and the results obtained from the quasiparticle model are systematically larger than those from the MIT bag model.
\begin{figure}[H]
	\centering
	\begin{minipage}{10cm}
		\includegraphics[width=10cm]{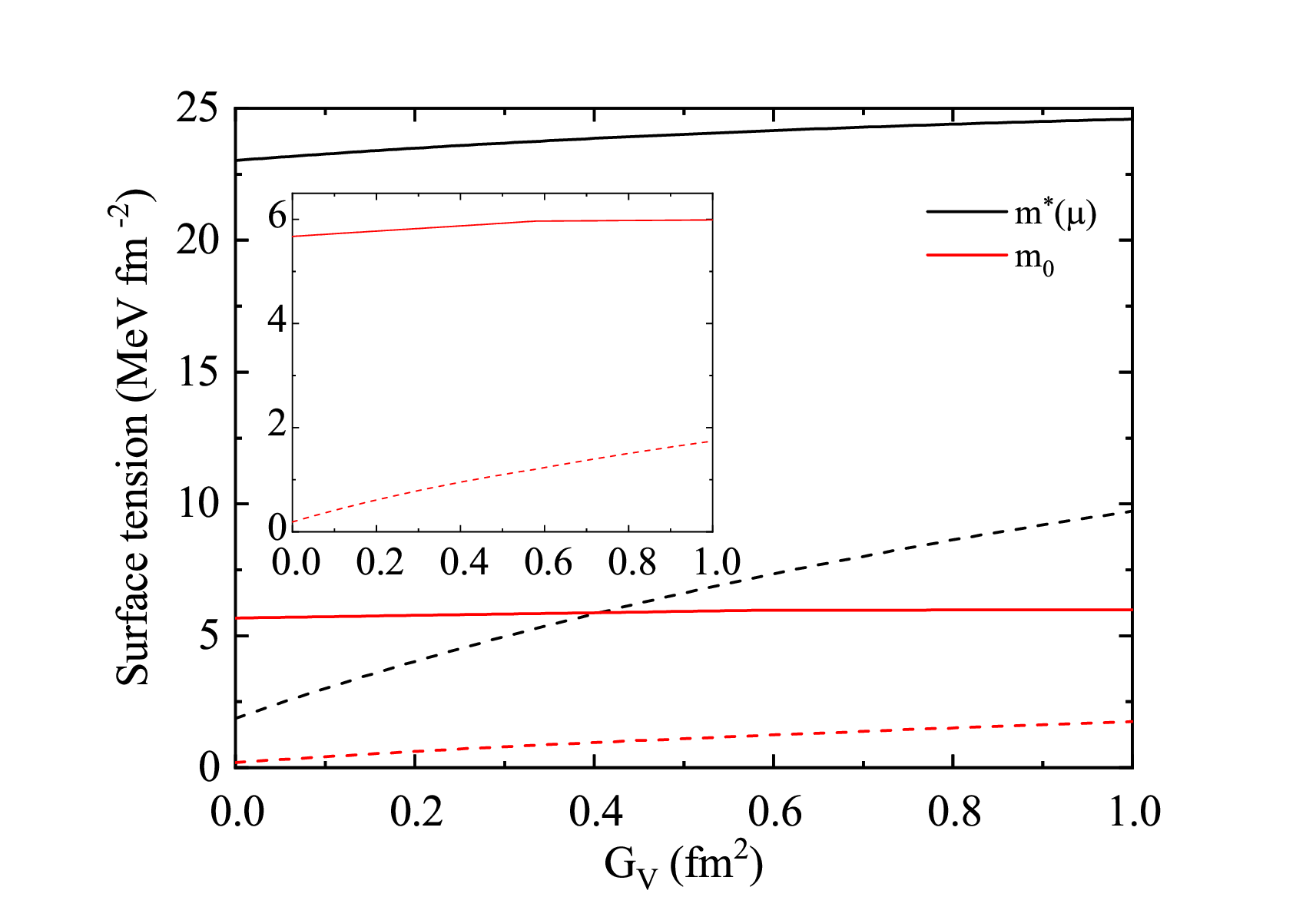}% Here is how to import EPS art
	\end{minipage}
	\caption{\label{fig:wide}Parallel (solid lines) and transverse (dashed lines) surface tensions as functions of the vector coupling constant ${G}_{V}$. }\label{fig:6}
\end{figure}

Fig. 7 shows only the transverse surface tension as a function of the magnetic field for a system with $g$=3, $R$=10 fm, and $\mu$=250 MeV. As shown in the figure, in the weak magnetic field region, the curves exhibit greater sensitivity to Landau level effects, manifesting oscillatory behavior. Under a much stronger magnetic field, the transverse surface tension without vector interaction (${G}_{V}$=0) would decrease and eventually vanish, whereas it increases with the magnetic field when vector interaction is taken into account. This difference stems from a key modification in the theoretical calculation (Eq. (\ref{eq:22})): without vector repulsive interaction, the transverse surface tension is determined solely by the contribution from the zeroth Landau level. However, when vector repulsive interaction is incorporated, the theoretical model includes a correction term that scales linearly with the magnetic field strength. The presence of this correction term leads to the observed enhancement of transverse surface tension under stronger magnetic fields. 
\begin{figure}[H]
	\centering
	\begin{minipage}{10cm}
		\includegraphics[width=10cm]{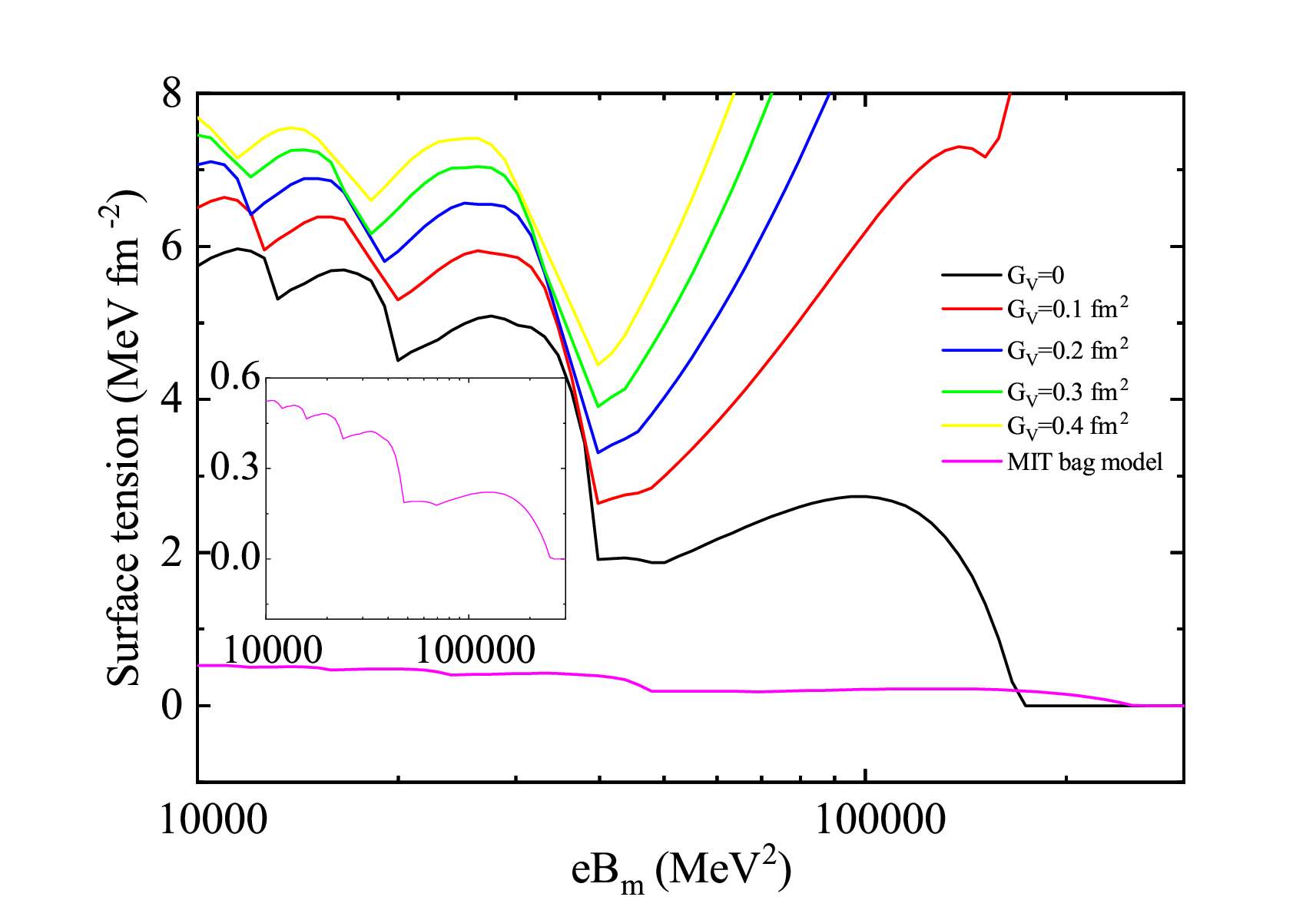}% Here is how to import EPS art
	\end{minipage}
	\caption{\label{fig:wide}The behavior of the transverse surface tension is
		shown as a function of the magnetic field at $g$=3, $R$=10 fm, and $\mu$=250 MeV. }\label{fig:7}
\end{figure}

\subsection{The stability of quark matter}

Edward Witten proposed that SQM might be absolutely stable at zero external pressure, potentially representing the true ground state of strongly interacting matter at finite baryon density\cite{Witten:1984rs}. In this section, we calculate the free energy per baryon for both two-flavor and three-flavor quark matter at zero temperature under a strong magnetic field. By comparing cases with and without vector interactions, we investigate their effects on matter stability. The free energy density without vector repulsive interactions is defined as ${\epsilon}_{\mathrm{wov}} =\Omega +\mu \textstyle\sum_{i}{n}_{i}$, whereas with vector interactions, it can be expressed as ${\epsilon}_{\mathrm{v}} ={\epsilon}_{\mathrm{wov}}\left ( {\mu }_{i}^{*}\right )+\frac{1}{2}{G}_{V}{\left ( \textstyle\sum_{i}{n}_{i}\right )}^{2}$\cite{Ju:2025mig}.

Fig. 8 shows the dependence of the free energy per baryon on the chemical potential for two-flavor and three-flavor quark matter under fixed parameters $g$= 4, $R$=10 fm, and $eB_m$= 0.1 GeV$^2$. The black, red, blue, green, and yellow curves correspond to the results obtained from the quasiparticle model with vector coupling constants ${G}_{V}$=0, 0.1, 0.2, 0.3, and 0.4 fm$^{2}$, respectively, while the pink curve represents the result from the MIT bag model with ${G}_{V}$=0. The dashed lines denote two-flavor quark matter, and the solid lines denote three-flavor quark matter. 

By comparing the quasiparticle model with the MIT bag model under the condition ${G}_{V}$=0, one finds that the free energy per baryon of two-flavor and three-flavor quark matter exhibits qualitatively consistent trends as a function of the chemical potential. Nevertheless, the numerical values obtained from the MIT bag model are systematically lower than those from the quasiparticle model.  In addition, the free energy per baryon of two-flavor quark matter is higher than that of three-flavor quark matter, indicating that three-flavor quark matter is more stable. The free energy per baryon for both two-flavor and three-flavor quark matter initially decreases and then increases with increasing chemical potential. The two curves coincide at lower chemical potentials because the strange quark does not contribute significantly in this regime. However, the minimum values of the free energy per baryon for both two-flavor and three-flavor quark matter are greater than the energy density of nuclear matter (930 MeV), which is generated by finite-volume effects. In Fig. 9, we present the results for a larger radius ($R$=100 fm), where the finite-volume effect becomes so insignificant that the minimum free energy per baryon for both cases is lower than 930 MeV. Furthermore, our calculations demonstrate that decreasing the radius leads to an increase in the free energy per baryon for both two-flavor and three-flavor quark matter.
\begin{figure}[H]
	\centering
	\begin{minipage}{10cm}
		\includegraphics[width=10cm]{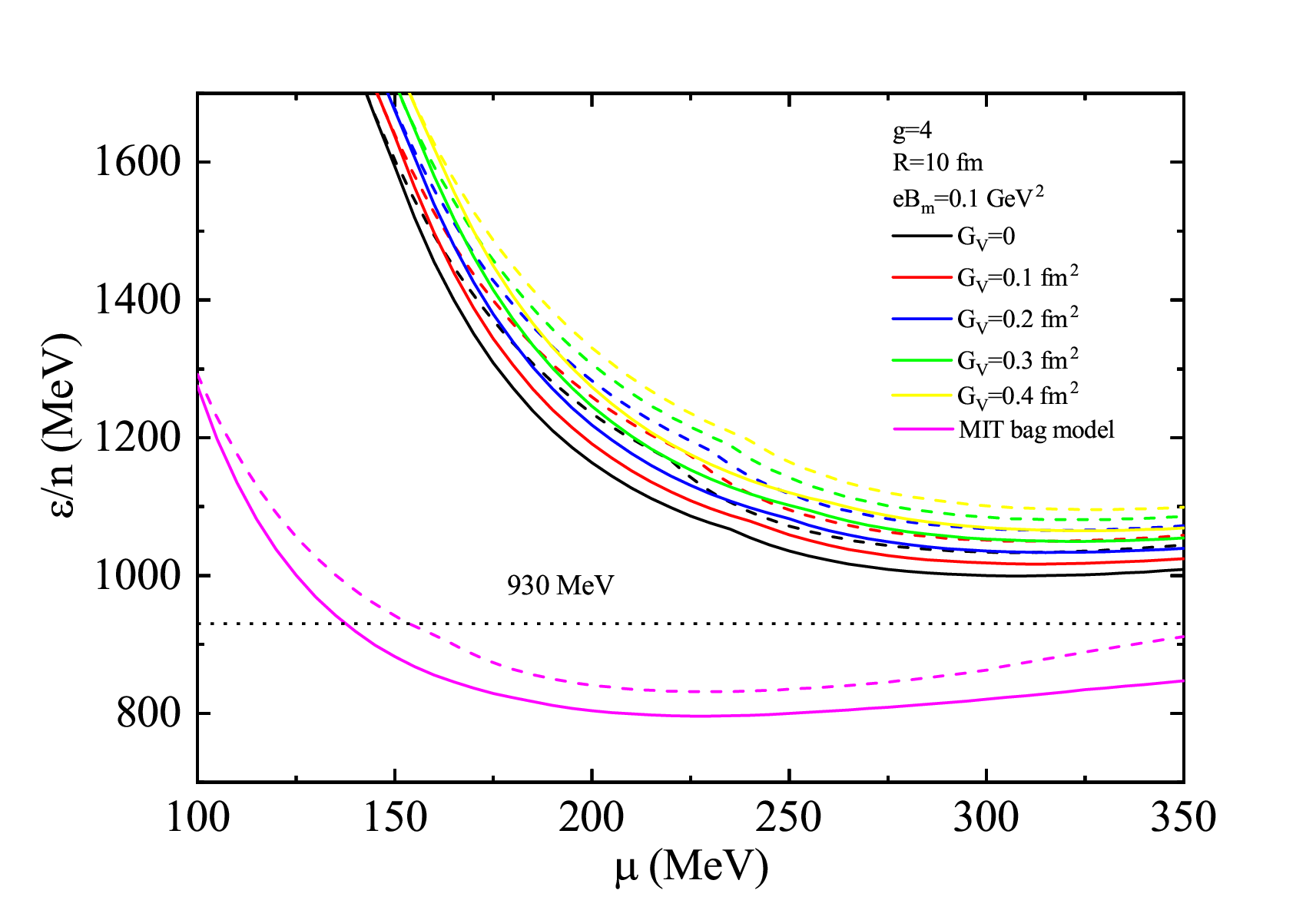}% Here is how to import EPS art
	\end{minipage}
	\caption{\label{fig:wide}The free energy per baryon as a function of the chemical potential for two-flavor and three-flavor quark matter at $g$= 4, $R$=10 fm, and $eB_m$= 0.1 GeV$^2$. The dashed and solid curves correspond to the two-flavor and three-flavor cases, respectively. }\label{fig:8}
\end{figure}
\begin{figure}[H]
	\centering
	\begin{minipage}{10cm}
		\includegraphics[width=10cm]{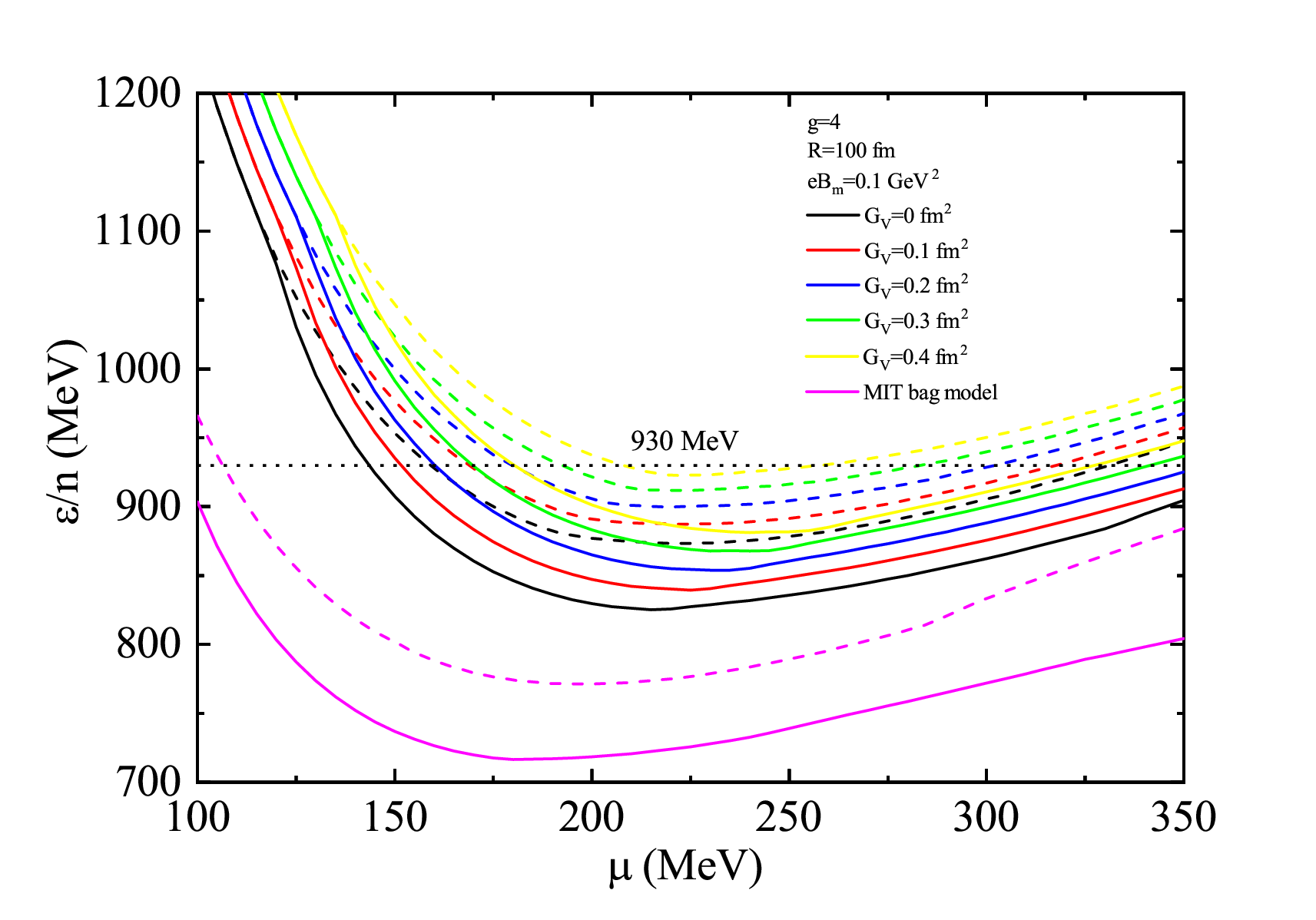}% Here is how to import EPS art
	\end{minipage}
	\caption{\label{fig:wide}At $g$= 4, $eB_m$= 0.1 GeV$^2$, and $R$=100 fm, the variation of free energy per baryon with chemical potential. }\label{fig:9}
\end{figure}
It can also be observed from Figs. 8 and 9 that the introduction of vector interactions increases the free energy per baryon for both two-flavor and three-flavor quark matter. This behavior indicates that vector repulsive interaction enhances the overall rigidity of the system, thereby increasing the energy cost required to excite the quark medium per baryon. To more intuitively illustrate the effect of vector coupling, Fig. 10 further presents the free energy per baryon for two-flavor and three-flavor quark matter as a function of the vector coupling constant under fixed parameters $g$= 4, $R$=10 fm, $\mu$=300 MeV, and $eB_m$= 0.1 GeV$^2$, for both the quasiparticle model and the MIT bag model. The dashed lines denote two-flavor quark matter, while the solid lines denote three-flavor quark matter. Regardless of the model employed, the free energy per baryon of both two-flavor and three-flavor quark matter increases monotonically with the vector coupling constant, indicating that enhanced vector interactions tend to reduce the stability of quark matter.
\begin{figure}[H]
	\centering
	\begin{minipage}{10cm}
		\includegraphics[width=10cm]{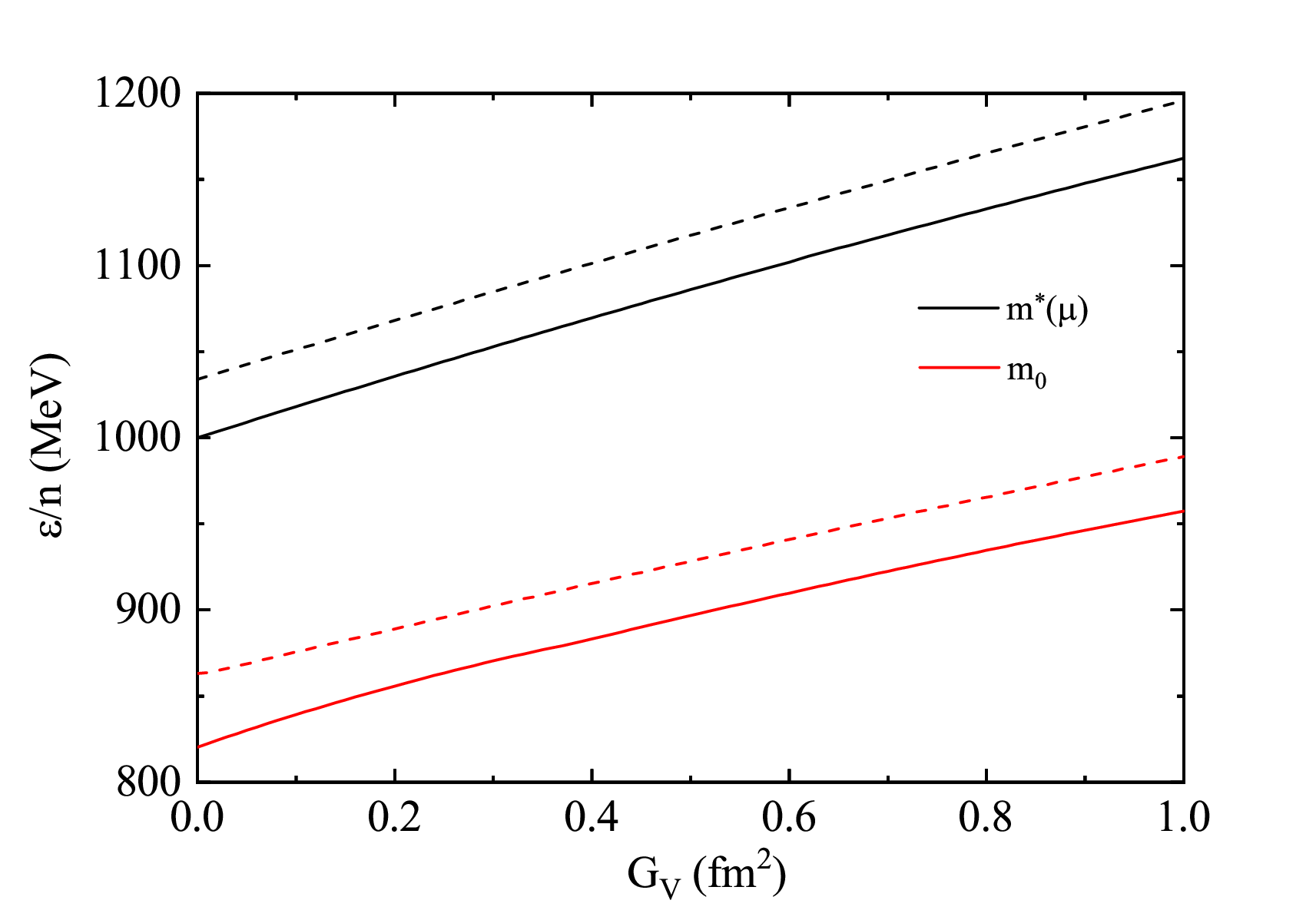}% Here is how to import EPS art
	\end{minipage}
	\caption{\label{fig:wide}Free energy per baryon versus the vector coupling constant ${G}_{V}$ for two-flavor (dashed line) and three-flavor (solid line) quark matter. }\label{fig:10}
\end{figure}
Next, we further investigate the behavior of the free energy per baryon for both two-flavor and three-flavor quark matter under varying magnetic fields at $g$=5, $R$=10 fm, and $\mu$=300 MeV in Fig. 11. 
By comparing the quasiparticle model with the MIT bag model under the condition ${G}_{V}$=0, one finds that for both two-flavor and three-flavor quark matter, the free energy per baryon decreases monotonically with increasing magnetic field strength, although the numerical values obtained from the MIT bag model remain systematically lower than those from the quasiparticle model. Moreover, at sufficiently strong magnetic fields, the free energy of three-flavor quark matter exceeds that of two-flavor quark matter. This behavior can be attributed to the combined effects of magnetic catalysis and the mass-charge competition mechanism, which dominate the properties of quark matter in strong magnetic fields. 
Further analysis reveals that strange quark matter satisfies the Witten stability hypothesis only when the magnetic field strength remains below 0.2 GeV$^2$. Notably, the free energy per baryon for both two-flavor and three-flavor quark matter increases monotonically with the vector coupling constant ${G}_{V}$. Consequently, for both two-flavor and three-flavor quark matter, the inclusion of vector interactions leads to an increase in the free energy per baryon, indicating that vector interactions reduce the stability of quark matter, which is in agreement with the previous opinion of Ref. \cite{Ju:2025mig}. 
\begin{figure}[H]
	\centering
	\begin{minipage}{10cm}
		\includegraphics[width=10cm]{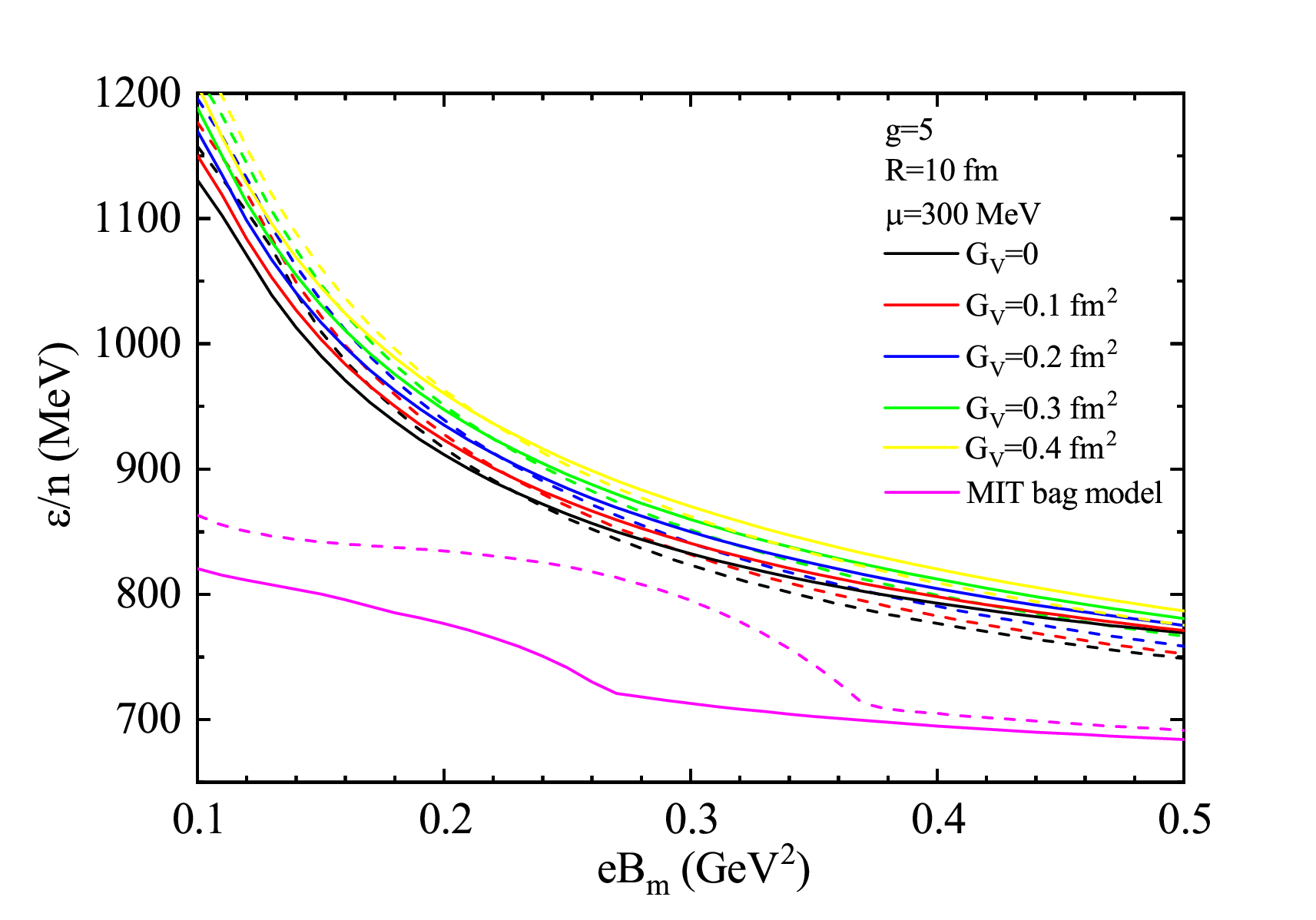}% Here is how to import EPS art
	\end{minipage}
	\caption{\label{fig:wide}The free energy per baryon as a function of magnetic field for two-flavor (dashed curves) and three-flavor (solid curves) quark matter at $g$= 5, $R$=10 fm, and $\mu$=300 MeV. }\label{fig:11}
\end{figure}

\section{Summary}
In this work, we investigate the anisotropic surface tension and free energy per baryon of quark matter within the quasiparticle model and the MIT bag model framework. By introducing a chemical-potential-dependent bag function to characterize medium effects, we find that the bag function decreases monotonically with increasing chemical potential. Notably, turning on vector interactions would result in a larger bag function. The finite-volume effects are treated by the multiple reflection expansion method, which reveals that the bag function diminishes with increasing system radius and asymptotically converges to the vacuum value $B_0$. It is shown that within a specific range of magnetic field intensities, the increase of magnetic fields would induce pronounced anisotropic surface tension. In particular, the parallel component increases monotonically with field strength, while the perpendicular component progressively decreases. 
Furthermore, the inclusion of vector interactions increases the surface tensions in both directions, and the surface tensions increase monotonically with the vector coupling constant. The curvature effect also contributes to the surface tension. Under a stronger magnetic field, the vector repulsive interaction leads to an increase in transverse surface tension with the magnetic field strength.

By comparing the free energy per baryon of two-flavor and three-flavor quark matter at different values of the vector coupling constant, we find that vector interactions tend to increase it, with larger coupling constants yielding higher values, thereby reducing the stability of quark matter. This is dramatically opposite to the magnetic field effect. Finally, we have shown strange quark matter is always consistent with the Witten hypothesis within a certain range of magnetic field intensities. The comprehensive theoretical framework developed in this work provides fundamental insights into understanding the phase structure and equation of state in the interiors of compact astrophysical objects, particularly neutron stars and quark stars.
\acknowledgments{ The authors would like to thank support from the
National Natural Science Foundation of China under the Grant Nos.
11875181, 12047571, and 11705163. This work was also sponsored by
the Fund for Shanxi "1331 Project" Key Subjects Construction.}
%%%%%%%%%%%%%%%%%%%%%%%%%%%%%%%%%%%%%%%%%%%%%%%%%%%%%%%%%%%%%%%%%%

\newpage 
\appendix
\renewcommand{\thesection}{\Alph{section}}  % 可选，默认已改
\section{Thermodynamic consistency with effective mass}
In the appendix sections, for the sake of convenience, we have disregarded the quark flavor indices in the variables.
In the grand canonical ensemble, the thermodynamic pressure of quarks with an effective quark mass $m^*(\mu)$ dependent on the chemical potential $\mu$ is modified by the background field $B^*(m^*)$ as:
\begin{eqnarray}
P(\mu,m^*)=-\Omega(\mu,m^*)-B^*(m^*),
\end{eqnarray}
where the thermodynamic potential $\Omega$ is from one flavor of quarks at zero temperature. It can be integrated up to the Fermi momentum $p_F=\sqrt{\mu^2-{m^*}^2}$ by the following integral,
\begin{equation}
\Omega=\frac{d}{2\pi^2}\int_0^{p_F}(E-\mu)p^2 dp,
\end{equation}
where $E=\sqrt{p^2+{m^*}^2}$ is the single particle energy function.

Since the pressure is expressed as a function of $\mu$, the chemical potential dependence of the quark mass will contribute to the total derivative of the pressure in an implicit part,
\begin{eqnarray}
\frac{d P}{d \mu}=-\frac{\partial \Omega}{\partial \mu}-\bigg(\underbrace{\frac{\partial \Omega}{\partial m^*} +\frac{\partial B^*}{\partial m^*}}_0\bigg)\frac{\partial m^*}{\partial \mu},\label{dPdu}
\end{eqnarray}
In the quasiparticle model, the effective mass is generally regarded as an order parameter determined by extremal conditions. The pressure should not change with the intermediate variable. The condition $\partial P/\partial m^*=0$ immediately leads to the terms in the parentheses of Eq. (\ref{dPdu}) being zero. Therefore the background field $B^*$ can be  regarded as a sum of medium term and vacuum term $B^*(\mu)=B_1+B_0$. The medium term is determined by the following integral,
\begin{equation}
B_1=-\int_{\mu_{\mathrm{IR}}}^\mu \frac{\partial \Omega}{\partial m^*} \frac{d m^*}{d \mu} d\mu,\label{B1}
\end{equation}
where the lower limit $\mu_{IR}$ is the root of the equation $\mu_{\mathrm{IR}}-m^*(\mu_{\mathrm{IR}})=0$. At the chemical potential $\mu_{\mathrm{IR}}$, one has $B_1=0$. These are only excitations of the underlying fields, but they are not ``real'' particles in the sense of being observable on-shell states. So the background field $B^*$ can return to the conventional bag constant $B_0$. At larger chemical potential in a dense medium, the confinement pressure is reduced by $B_1$. Thermodynamic consistency can be demonstrated by the fact that the zero-pressure point is consistent with the minimum value of the free energy.

Let us evaluate the first term in Eq.(\ref{dPdu}).
In the integration of $\Omega$, the upper limit $p_F$ depends on the effective mass $m^*$, so we have the derivative,
\begin{eqnarray}
\frac{\partial \Omega}{\partial \mu}=-\frac{d }{6\pi^2}p_F^3-\frac{d}{2\pi^2}\bigg\{  \underbrace{[E(p_F)-\mu]}_0  \frac{dp_F}{d\mu} \bigg\},
\end{eqnarray}
where the term in the brace is from the derivative of the upper limit of the integral multiplied by the value of the integrand at that upper limit, and it vanishes due to the definition of Fermi energy. So the $\mu$-dependent terms $B^*$ and  the boundary of the Fermi surface will not contribute apparently in the derivative of the pressure. The quark number density $n$ is still expressed in the conventional formula,
\begin{equation}
n=\frac{d }{6\pi^2}p_F^3.
\end{equation}
The background field certainly contributes to the total energy density. Therefore the energy density satisfies the basic thermodynamic relation
\begin{equation}
\varepsilon=-P+\mu n.
\end{equation}
It can be concluded that the number density and energy density are of the same form as that of a Fermi gas with normal particle mass.
\section{Medium-dependent MRE formulation}

By employing the MRE formulation, one can investigate matter of finite size. For the spherical volume with radius $R$, the thermodynamic pressure is modified by the density of states as,
\begin{equation}
P(\mu,R,m^*)=-\Omega(\mu, m^*,\rho_{\mathrm{MRE}})-B(m^*). \label{pressure}
\end{equation}
The density of states of degererate quarks obeying confining boundary condition with its momentum $p$ is expressed as

\begin{eqnarray}
\rho_{\mathrm{MRE}}(\mu,p)&=&1+\frac{6\pi ^{2} }{pR}f_{S}+\frac{12\pi ^{2}}{\left ( pR\right )^{2}} f_{C}, \\
f_{S}(\mu,p)&=&-\frac{1}{8\pi }\left( 1-\frac{2}{\pi }\arctan\frac{p}{m^{*} }\right ), \\
f_{C}(\mu,p)&=&\frac{1}{12\pi ^{2}}\left [ 1-\frac{3p}{2m^{*}}\left ( \frac{\pi }{2}-\arctan \frac{p}{m^{*}}\right )\right ].
\end{eqnarray}

By promoting the mass to a medium-dependent quantity, the effective mass is usually larger than the current mass. Compared with the standard MRE formulation with fixed constant mass, the larger effective mass will suppress the density of states $\rho_{\mathrm{MRE}}$ in momentum space. Fortunately, the additional term $B$ will compensate the effect. This can be understood by the total derivative of the pressure with the chemical potential,
\begin{equation}
\frac{d P}{d\mu}=-\frac{\partial \Omega}{\partial \mu} -\bigg(\underbrace{\frac{\partial \Omega}{\partial m^*} +\frac{\partial \Omega}{\partial \rho_{\mathrm{MRE}}}\frac{\partial \rho_{\mathrm{MRE}}}{\partial m^*} +\frac{\partial B^*}{\partial m^*}}_0\bigg) \frac{d m^*}{d \mu}.  
\end{equation}
Similar to Eq. (\ref{dPdu}), $\partial P/\partial m^*=0$ leads to the term in the parenthesis being zero. Then we can obtain the bag function $B(\mu)$,
\begin{equation}
B^*(\mu)=B_1+B_2+B_0.
\end{equation}
The $B_1$ term is the same as the result of bulk matter in Eq. (\ref{B1}). The term $B_2$ due to the $\mu$-dependence of the density of states is
\begin{equation}
B_2=-\int_{\mu_{\mathrm{IR}}}^\mu \frac{\partial \Omega}{\partial \rho_{\mathrm{MRE}}}\frac{\partial \rho_{\mathrm{MRE}}}{\partial m^*}  \frac{d m^*}{d \mu} d\mu .
\end{equation}
The remaining derivative of $\Omega$ with respect the chemical potential can be evaluated as following,
\begin{eqnarray}
\frac{\partial \Omega}{\partial \mu}&=&-\frac{d}{2\pi^2}\Bigg\{ \int_{\Lambda_{\mathrm{IR}}}^{p_F}\rho_{\mathrm{MRE}}(\mu,p)dp+\underbrace{[E(p_F)-\mu]}_0 \rho_{\mathrm{MRE}}(\mu,p_F) \frac{dp_F}{d\mu}\nonumber \\
&& \phantom{-\frac{d_i|q_i B_m|}{2\pi^2}} - [E(\Lambda_{\mathrm{IR}})-\mu] \underbrace{\rho_{\mathrm{MRE}}(\mu,\Lambda_{\mathrm{IR}})}_0\frac{d\Lambda_{\mathrm{IR}}}{d\mu} \Bigg\},
\end{eqnarray}
The second term is zero due to the same Fermi energy as in bulk matter. The third term corresponds to the lower limit, which is zero by virtue of the infrared condition. So the quark number density $n$ is in the conventional expression as expected,
\begin{equation}
n=\frac{d P}{d \mu}=-\frac{\partial \Omega}{\partial \mu}.
\end{equation}
This is a direct consequence of the thermodynamic self-consistency. It will not be changed by the effective mass in the MRE formulation.

In conclusion, the density of states should be modified by employing the effective quark mass. From the pressure function (\ref{pressure}), the modified density of states $\rho'_{\mathrm{MRE}}$ can be formally regarded as,
\begin{equation}
\rho'_{\mathrm{MRE}}=\rho_{\mathrm{MRE}}+\int_{\mu_{\mathrm{IR}}}^\mu \frac{\partial \rho_{\mathrm{MRE}}}{\partial m^*}  \frac{d m^*}{d \mu} d\mu ,.
\end{equation}
The additional term compensates the supression of original $\rho_{\mathrm{MRE}}$. The derivative of the density of states with respect to the effective mass is
\begin{equation}
\frac{\partial \rho_{\mathrm{MRE}}}{\partial m^*}=-\frac{3}{2R}\frac{1}{E^2}+ \frac{3}{2 R^2 p^3}[-\frac{1}{x(1+x^2)}+\frac{\arctan(x)}{x^2} ],
\end{equation}
with the dimensionless parameter $x=m^*/p$.
\end{document}